ARTICLE    OPEN

# Squeezing-enhanced Raman spectroscopy

Yoad Michael[1], Leon Bello[1], Michael Rosenbluh[1] and Avi Pe'er[1]*

The sensitivity of classical Raman spectroscopy methods, such as coherent anti-stokes Raman spectroscopy (CARS) or stimulated Raman spectroscopy (SRS), is ultimately limited by shot-noise from the stimulating fields. We present the complete theoretical analysis of a squeezing-enhanced version of Raman spectroscopy that overcomes the shot-noise limit of sensitivity with enhancement of the Raman signal and inherent background suppression, while remaining fully compatible with standard Raman spectroscopy methods. By incorporating the Raman sample between two phase-sensitive parametric amplifiers that squeeze the light along orthogonal quadrature axes, the typical intensity measurement of the Raman response is converted into a quantum-limited, super-sensitive estimation of phase. The resonant Raman response in the sample induces a phase shift to signal-idler frequency-pairs within the fingerprint spectrum of the molecule, resulting in amplification of the resonant Raman signal by the squeezing factor of the parametric amplifiers, whereas the non-resonant background is annihilated by destructive interference. Seeding the interferometer with classical coherent light stimulates the Raman signal further without increasing the background, effectively forming squeezing-enhanced versions of CARS and SRS, where the quantum enhancement is achieved on top of the classical stimulation.

## INTRODUCTION

Quantum-enhanced measurements utilize the unique correlation properties of non-classical light for highly sensitive detection. Common examples include NOON[1] and squeezing-based[2,3] interferometers that employ entangled quantum states to achieve sub-shot-noise phase sensitivity. This enhancement can be useful for measurements of extremely weak signals, with a crowning example being the detection of gravitational waves.[4,5] A major field that can greatly benefit from sub-shot-noise detection is Raman spectroscopy, which is widely used for chemical sensing,[6–8] due to its ability to identify the molecular contents of a sample based on its Raman fingerprint spectrum. Raman spectroscopy is, therefore, an ideal contrasting method for chemically-resolved microscopy[9] with no prior preparation or fluorescent tagging of the target molecule required. However, the major challenge for Raman sensing is the relative weakness of the Raman response, which is orders of magnitude weaker than fluorescence, and may often be obscured by the shot-noise of other stimulated light-matter interactions.

In coherent anti-stokes Raman spectroscopy (CARS),[10] a Raman sample is excited by a strong pump wave (frequency $\omega_p$) and a Stokes wave (idler, frequency $\omega_i$) that interact within the sample to generate an anti-Stokes (signal) wave at frequency $\omega_s = 2\omega_p - \omega_i$ via four-wave mixing (FWM). When the frequency difference between the pump and Stokes field matches a molecular vibration/rotation in the sample, the generated anti-Stokes field is resonantly enhanced, indicating that the Raman frequency shift of the signal (with respect to the pump) acts as a molecular fingerprint. However, since FWM is a parametric process,[11] non-resonant FWM can also occur via virtual levels, resulting in a non-resonant background that is not chemically specific.[12] In diluted samples, where the target molecule is surrounded by large quantities of background molecules (e.g., a protein dissolved in water within a biological cell), the non-resonant background from the environment (water) can become a major limiting factor to the sensitivity of measurement, since it dominates over and obscures the weak resonant Raman signal from the target molecule (protein). The fundamental limit to the sensitivity of standard CARS is, therefore, the noise associated with the non-resonant background, indicating that suppression of this background is a major goal for CARS spectroscopy, and several methods have been proposed in past research to address it: Pulse shaping was applied to reduce the peak power of the exciting pulses[13] (and hence the non-resonant background), epi-CARS that detects only the back-scattered Raman signal[14] (which is primarily resonant), and polarization CARS that rejects the non-resonant signal based on polarization.[15] All these methods rely on some specific property of the sample/light to suppress the background, and even with ideal suppression, all classical detection methods are ultimately limited by shot-noise. In stimulated Raman spectroscopy (SRS),[16] only the resonant Raman response is observed, through sensitive measurement of the weak nonlinear gain that the Stokes field experiences in the presence of a strong pump. While SRS is free of the non-resonant background, its sensitivity is fundamentally limited by the shot-noise of the coherent Stokes seed.

Here, we propose and present the theoretical analysis of a sample-independent configuration for measurement of the resonant Raman response beyond the shot-noise limit, by recasting the typical measurement of the Raman intensity into a quantum-enhanced estimation of the nonlinear Raman phase. In our proposal, we place the Raman sample between two phase-sensitive optical parametric amplifiers (OPAs)—forming a non-linear version of the well-known SU(1,1) interferometer (Fig. 1a), which employs the squeezing of the OPAs[17,18] to measure an induced optical phase shift beyond the shot-noise limit (see a brief overview of the SU(1,1) interferometer in the Methods and comprehensive reviews in ref. [3,19]). By probing the Raman interaction in the sample with broadband two-mode squeezed light (see Fig. 1b), generated via FWM in the OPAs,[20] the resonant

[1]Department of Physics and BINA Center for Nanotechnology, Bar-Ilan University, Ramat Gan 5290002, Israel. *email: Avi.peer@biu.ac.il

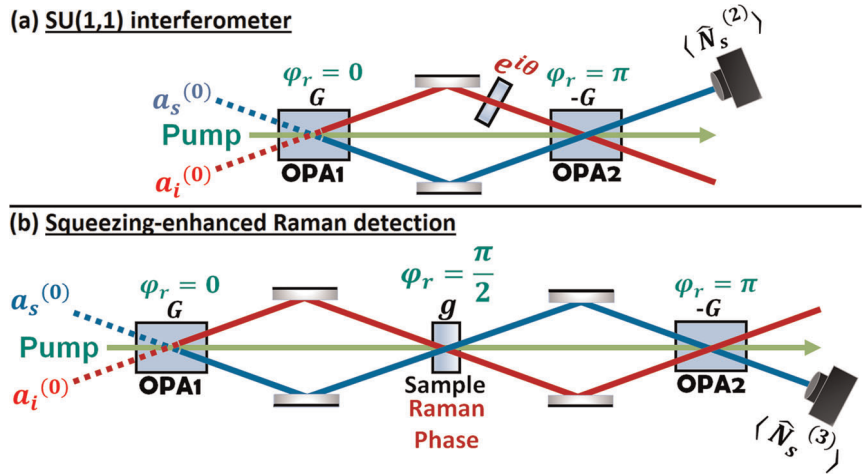

**Fig. 1** Squeezing-enhanced Raman detection in an SU(1,1) interferometer. **a** The standard SU(1,1) interferometric detection of a linear phase: Two phase-sensitive OPAs of equal gain and opposite pump phases are arranged in series (balanced configuration). OPA1 amplifies signal-idler pairs at the input (vacuum), while OPA2 is shifted in pump phase to exactly reverse the amplification of OPA1 and return the light back to its original input state. When a linear phase shift $\theta$ is introduced, the cross-cancellation of the amplifiers is no longer complete, and additional light is detected at the output. This light directly corresponds to the induced phase shift, which is detected with sub-shot-noise sensitivity due to the squeezing of the OPAs. **b** Our scheme for squeezing-enhanced Raman detection: The Raman sample is introduced into a balanced SU(1,1) interferometer. The sample acts as a phase-sensitive parametric amplifier that induces a phase shift into the interferometer, and therefore generates some light at the (previously) dark output, as explained in the text

signal of the sample is enhanced by the squeezing ratio of the OPAs, while the non-resonant background is completely eliminated by destructive interference due to the inherent $\pi/2$ difference in the phase response between the resonant and non-resonant interactions. This configuration effectively forms a quantum-enhanced version of interferometric CARS.[21,22]

Our theoretical analysis of this squeezing-enhanced Raman spectroscopy method is organized as follows: Section "Squeezing-enhancement of spontaneous Raman spectroscopy" highlights the squeezing enhancement of the resonant Raman signal by analysis of a resonant Raman sample placed inside a lossless SU(1,1) interferometer. Section "Complete suppression of the non-resonant background" analyses the additional effect of a non-resonant interaction in the Raman sample, by distinguishing between the non-resonant background and the resonant signal, which yields complete elimination of the background due to the inherent $\pm\pi/2$ phase difference between the resonant and non-resonant interactions. Section "Coherent seeding of the interferometer: squeezing-enhanced CARS and SRS" introduces coherent seeding of the interferometer by classical fields, showing that the squeezing enhancement of section "Squeezing-enhancement of spontaneous Raman spectroscopy" can be added on top of the classical stimulation of the Raman interaction. This squeezing-enhancement can be applied to any classical Raman method, such as CARS and SRS. Finally, section "Detection sensitivity in the presence of loss" incorporates loss into the analysis, both internal and external to the interferometer, which is a critical consideration for experimental realizations of any squeezing application, showing that the scheme maintains its sub-shot-noise sensitivity even with practical levels of loss, as long as the phase of OPA2 is appropriately tuned.

## RESULTS

### Squeezing-enhancement of spontaneous Raman spectroscopy

The Raman sample, which can be considered as a weak parametric amplifier via Raman-based FWM, is placed inside a nonlinear interferometer that is composed of two external OPAs (see Fig. 2a for illustration), where each OPA amplifies one quadrature (and attenuates the other quadrature) of the two-mode signal-idler field. Using the field operators of the signal and idler $\hat{a}_{s,i}$, we can define the quadratures of the two-mode field as: $\hat{X} = \hat{a}_s + \hat{a}_i^\dagger$ and $\hat{Y} = i(\hat{a}_s^\dagger - \hat{a}_i)$ (See Methods and ref. [23,24] for additional information). The two-mode quadratures represent the cosine and sine oscillation components of the combined signal-idler field and form a pair of conjugate variables, which obey the standard commutation relation $[\hat{X}, \hat{Y}] = 2i$. As previously stated, an OPA amplifies one quadrature $\hat{X} \to \hat{X}e^G$ and attenuates the other $\hat{Y} \to \hat{Y}e^{-G}$, and the quadratures evolve as shown in Fig. 2 for each amplifier. The two OPAs are arranged in a "crossed" configuration, where the quadrature attenuation axis of OPA2 matches the amplification axis OPA1 and vice-versa, such that OPA2 exactly reverses the squeezing of OPA1 (equal gain, see Fig. 2b). The orientation of the amplification/attenuation axes of the OPAs (and the Raman sample) is determined by the relative phase $\phi_r = 2\phi_p - (\phi_i + \phi_s)$ between the pump and signal-idler fields. This relative phase is a tunable parameter that can be controlled by adjusting the phase of either the pump, signal, or idler (for example, by adjusting the path of the beams). The squeezed signal-idler pairs generated in OPA1 (with a relative phase $\phi_r = 0$) interact parametrically with the Raman sample at an intermediate relative phase $\phi_r = \pi/2$. In practice, this relative phase could be achieved by changing the optical path of the beams, but if OPA1 and OPA2 generate non-resonant light (for example, by using Photonic Crystal Fibers[25]), then the resonant Raman amplification of the sample will occur at $\phi_r = \pi/2$ with no additional phase adjustments required (see section "Complete suppression of the non-resonant background" hereon for further discussions about the resonant and non-resonant interactions). This relative phase is translated to an amplification axis of 45° with respect to OPA1, which in-turn rotates the squeezing ellipse. OPA2 then amplifies at an opposite relative phase $\phi_r = \pi$, which without a Raman sample will completely reconvert the signal-idler pairs back to pump light (leaving a vacuum output). In the presence of a Raman sample, the cancellation is incomplete, elevating the signal intensity at the output above zero (from its initial vacuum state).

Let us first calculate the light intensity (number of photons) at the output of this squeezing-enhanced Raman configuration. When passing through the amplifiers (and the sample), we can use the input-output relations of the field operators at each

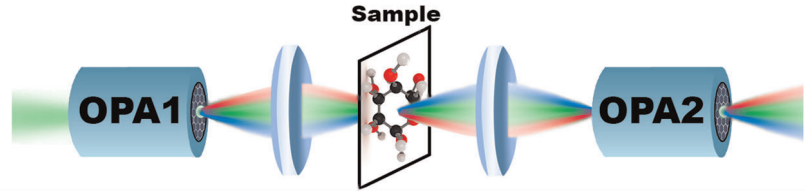

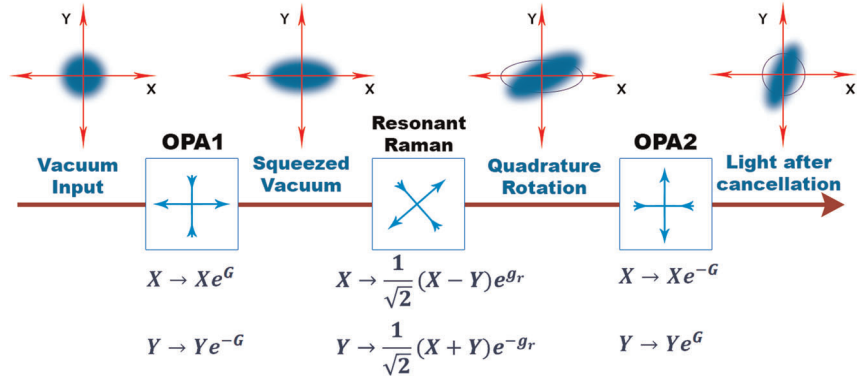

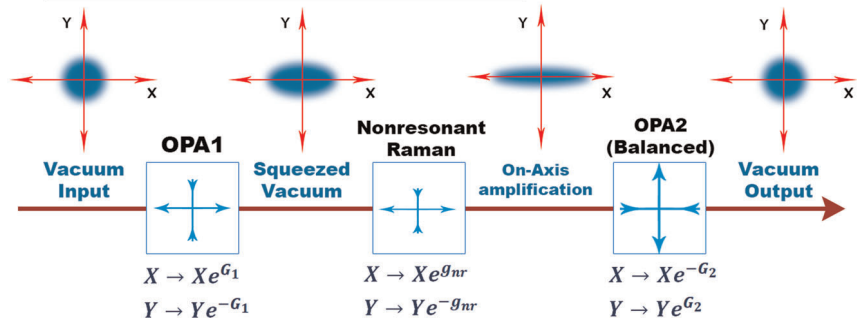

**Fig. 2** Overview of the signal-idler two-mode quadratures within the squeezing-enhanced Raman configuration. **a** Conceptual illustration of the scheme, where photonic crystal fibers represent the OPAs. **b**, **c** Show the two-mode quadrature dynamics of the light throughout the different amplifiers for resonant and non-resonant interactions respectively. The top line of graphs in **b**, **c** portrays the quadrature map of the two-mode light at various locations along the interferometer, and the bottom part shows the amplification/attenuation axes of each amplifier (indicated by arrows pointing outwards/inwards). OPA1 and OPA2 are set in a crossed (orthogonal) configuration, and are assumed to be ideal squeezers. In the resonant case **b** the Raman sample is set to a relative phase of $\phi_r = \pi/2$ (45° amplification axis), which rotates the squeezing-ellipse of the two-mode light, resulting in non-vacuum output. The transparent ellipses/circles illustrate the state of the light in the absence of a sample, where the output returns exactly to the input state (vacuum). **c** For a non-resonant Raman sample, the relative phase is $\phi_r = 0$, indicating that the amplification axis of the non-resonant sample is aligned with that of OPA1, and the two-mode ellipse is not rotated, but rather further squeezed in the same direction. The non-resonant gain $g_{nr}$ unbalances the interferometer by effectively increasing/decreasing the gain of OPA1 $G_1$, which can be negated by tuning the gain of OPA2 to $G_2 = G_1 + g_{nr}$

amplifier:[24]

$$\hat{a}_{s,i}^{(1)} = A\hat{a}_{s,i}^{(0)} + B\hat{a}_{i,s}^{\dagger(0)}, \quad (1)$$

$$\hat{a}_{s,i}^{(2)} = C\hat{a}_{s,i}^{(1)} e^{i\pi/4} + D\hat{a}_{i,s}^{\dagger(1)} e^{-i\pi/4}, \quad (2)$$

$$\hat{a}_{s,i}^{(3)} = A\hat{a}_{s,i}^{(2)} e^{i\pi/4} + B\hat{a}_{i,s}^{\dagger(2)} e^{-i\pi/4}, \quad (3)$$

where $A$ and $B$ represent the operation of the OPAs: $A = \cosh(G)$ and $B = \sinh(G)$, with $G$ the gain of the amplifiers, and $C$ and $D$ are associated with the Raman sample, $C = \cosh(g_r)$ and $D = \sinh(g_r)$ with $g_r$ the gain of the Raman sample (assumed for now to be purely resonant and narrowband). The relative phase between the pump, signal and idler varies throughout the different parametric amplifiers, and is adjusted such that the two external OPAs are orthogonal to each other, with the Raman sample set to an intermediate phase of $\phi_r = \pi/2$. Thus, we apply a $\pi/4$ phase to both the signal and idler fields (accumulating to $\phi_r = \pi/2$) twice:

first after OPA1, which rotates the quadrature amplification axis of the Raman sample to 45°, and a second time after the sample, which sets OPA2 orthogonal to OPA1. The total accumulated relative phase for the FWM light before OPA2 is, therefore, $\phi_r = \pi$, indicating that in the absence of a Raman sample ($C = 1$, $D = 0$), the scheme is reduced to the standard SU(1,1) interferometer with complete destructive interference. With the Raman sample added, a nonlinear phase shift is induced by the resonant response of the Raman sample, and photon number at the signal output is (see Methods for derivation):

$$\begin{aligned}\left\langle \hat{N}_s^{(3)} \right\rangle &= |D|^2(|A|^2 + |B|^2 + 2|A||B|) \\ &= \sinh^2(g_r)\cosh^2(2G).\end{aligned} \quad (4)$$

Equation (4) shows the signal at the output of the interferometer, and highlights the increase in the overall signal due to the stimulation of the Raman sample by the two-mode squeezed light generated in OPA1, whose noise is then eliminated by OPA2. We can compare the result of Eq. (4) to that of spontaneous Raman emission: $\langle \hat{N}_s^{(3)} \rangle = \sinh^2(g_r)$, which corresponds to setting the OPA gain to zero $G = 0$. Thus, Eq. (4) illustrates the coupling between the parameters of the sample ($C$, $D$, $g_r$) and the parameters of the crossed amplifiers ($A$, $B$, $G$), creating an effective squeezing-enhanced signal, but with no added background (the non-resonant background can be eliminated by tuning the gain of the external OPAs, as we explain in the next section). Thus, although the sample is stimulated by the FWM light of OPA1, the interferometer *conceals* this stimulation completely, acting as an effective black-box, which to an external observer appears just like a spontaneous Raman scatterer, but with a higher effective (squeezing-enhanced) signal. An analogy from the field of interferometry can be given here: In standard Mach-Zender SU (2) interferometers, the sensitivity of the interferometer is limited by shot-noise, but can be improved by seeding the unused port of the interferometer with squeezed light, resulting in "squeezing-enhanced" estimation of phase. In the scheme proposed here, we place the Raman sample between two-mode quadrature-squeezing OPAs: OPA1 generates squeezed light that stimulates the Raman interaction in the sample, while OPA2 attempts to unsqueeze the light back into vacuum. The overall result is a "squeezing-enhanced" detection of the Raman signal. As will be explained in section "Coherent seeding of the interferometer: squeezing-enhanced CARS and SRS", this effective Raman black-box remains applicable also for stimulated Raman techniques, such as CARS and SRS, by seeding the interferometer with a coherent idler (rather than vacuum).

It should also be noted that due to dispersion, the Raman sample will also induce a linear phase shift which is not Raman-specific. Control over this phase is possible using other dispersive media, as shown in ref. [26,27] effectively negating it.

### Complete suppression of the non-resonant background

In the previous section, we considered only the contribution of the resonant Raman process to the output intensity. The Raman response of the sample is generally complex with respect to the pump drive, and its phase varies spectrally across the resonance from $\phi_r = 0$ below resonance, through $\phi_r = \pi/2$ on resonance to $\phi_r = \pi$ above resonance[28] (not to be confused with the phases of the interferometer). The imaginary part of the Raman response is associated with the Raman absorption/gain, which is maximal on resonance, whereas the real part of the response is associated with dispersion, which is nulled on resonance, just like a driven two-level system.[29] Thus, we should treat the Raman sample as a parametric amplifier capable of two separate amplifications simultaneously: First, the resonant, phase-shifting amplification at $\phi_r = \pi/2$ (discussed thus far) and an additional non-resonant, non-phase-shifting amplification at $\phi_r = 0$ (or $\phi_r = \pi$ above resonance). In the two-mode quadrature representation, this amplification occurs "on-axis" with the amplification axis of OPA1, and acts as direct extension to the gain of OPA1 (see Fig. 2c)—it amplifies the same quadrature (the same relative phase) as OPA1, indicating that the non-resonant contribution can be completely nulled by varying the gain of the OPAs. This concept is similar to the experiments of Lupke[30] and Lee,[31] where the non-resonant background of the sample was canceled by placing an additional, non-Raman FWM medium, whose non-resonant background destructively interfered with the background from the sample. The scheme proposed here benefits from the background cancellation in combination with the inherent squeezing of the external OPAs to enhance the Raman signal.

To observe this background cancellation mathematically, let us first consider a purely non-resonant sample: after passing through OPA1 (with gain $G_1$), the sample performs non-resonant amplification with gain $g_{nr}$ and at a relative phase $\phi_r = 0$. This phase is identical to that of OPA1, which effectively extends the same amplification process, and in the quadrature picture, performs on-axis amplification (see Fig. 2c). OPA2 is set as before to negate the amplification of OPA1, this time with different gain $G_2 \neq G_1$. The number of photons at the output is then:

$$\langle \hat{N}_s^{(3)} \rangle_{nr} = \sinh^2(G_1 + g_{nr} - G_2). \quad (5)$$

Setting the gain of OPA2 to $G_2 = G_1 + g_{nr}$ will null the non-resonant output completely.

In practice, real Raman samples will have both resonant Raman gain from the molecule of interest and non-resonant gain from background molecules (e.g., solvent). For simplicity, we assume here that the Raman sample is composed of a target substance that produces a resonant Raman signal, surrounded by a uniform medium whose vibrational resonances do not overlap with that of the target molecule, such that it produces a non-resonant background (in our example, below the resonance). We can think of the sample as a mixture of many infinitesimal parametric amplifiers that perform either resonant amplification at $\phi_r = \pi/2$ or non-resonant amplification at $\phi_r = 0$ with some unknown ordering. The resonant and non-resonant gains obviously do not commute, but since the gain of the entire sample is small compared to the OPAs ($g_{nr}, g_r \ll G$), it is fair to neglect the cross-interaction between the resonant and non-resonant gains within the sample ($\sinh^2(g_r)\sinh^2(g_{nr})$) compared to the interaction of the sample with the external OPAs. Specifically, we can see from Eq. (5) that the two extreme possibilities of ordering the gain in the sample - either setting the non-resonant gain entirely before the resonant gain (adding slightly to the gain of OPA1 $G_1 \rightarrow G_1 + g_{nr}$) or entirely after the resonant gain (decreasing slightly $G_2 \rightarrow G_2 - g_{nr}$), both lead to the same gain balancing outcome—increase $G_2$ to compensate for $g_{nr}$. Thus, the effect of ordering within the sample must be of higher order, and we may neglect it for weak samples. We, therefore, treat the sample as two separate parametric amplifiers (resonant and non-resonant) that are placed in series, which yields the following photon number at the output:

$$\langle \hat{N}_s^{(4)} \rangle = \sinh^2(g_r)\cosh^2(2G_2). \quad (6)$$

The result of Eq. (6) remains similar to Eq. (4) even in the presence of the non-resonant background, which is now fully suppressed, indicating that the interferometric scheme still behaves as a spontaneous squeezing-enhanced Raman black-box. Based on this result, we can assume in the upcoming sections that the non-resonant contribution is canceled by gain balancing. Additional information about the unbalanced interferometer is provided in the Methods.

### Coherent seeding of the interferometer: squeezing-enhanced CARS and SRS

Let us now examine the CARS response of our squeezing-enhanced Raman configuration from section "Squeezing-enhancement of spontaneous Raman spectroscopy" by subjecting it to a strong coherent idler input.[32,33] Note first that by simple extension of the treatment in section "Complete suppression of the non-resonant background", we may assume that the non-resonant background is suppressed by gain balancing of the external OPAs, and consider only the resonant response (the stimulation affects both the OPAs and the sample in the same manner, leaving the gain balancing unchanged). This assumption holds true regardless of the input state of the signal or idler. The output photon number of the signal $\langle \hat{N}_s^{(3)} \rangle$ for the seeded configuration $|0_s, a_i\rangle$ (input of vacuum signal, strong coherent state idler) is then (see Methods for derivation)

$$\langle \hat{N}_s^{(3)} \rangle = (1 + |a_i|^2)\cosh^2(2G)\sinh^2(g_r), \quad (7)$$

which is directly equivalent to the expression of standard CARS with the additional enhancement of the Raman signal due to the squeezing inside the interferometer, and with full inherent suppression of the non-resonant background.

To consider our scheme for SRS, let us examine the increase in intensity (photon number) of the idler beam:

$$\langle \hat{N}_i^{(3)} \rangle - |a_i|^2 = \tfrac{1}{2}(1+|a_i|^2)(1+\cosh(4G))\sinh^2(g_r) \qquad (8)$$
$$= (1+|a_i|^2)\cosh^2(2G)\sinh^2(g_r).$$

In contrast to CARS, SRS requires lock-in detection to separate the Raman signal from the coherent idler input, but in principle, both methods benefit from the squeezing-enhancement of the signal due to the OPAs.

Consequently, the 'Raman black-box' concept, which claims that the interferometer can behave exactly like a normal Raman sample but with an enhanced signal due to the squeezing, extends also to stimulated interactions. The squeezing effect on the sample appears only internally between the crossed OPAs, and an external observer will not be able to differentiate the ideal interferometer configuration from a simple, high-signal resonant Raman sample (unless he can 'look inside the box').

Detection sensitivity in the presence of loss
Let us now evaluate the sensitivity of measurement for our suggested scheme under practical conditions by calculating the minimum detectable resonant Raman gain $(g_{min}^2)$ of the sample in the presence of photon-loss. This sensitivity can be calculated by error-propagation analysis:

$$g_{min}^2 = \frac{\langle \hat{N}_s^2 \rangle - \langle \hat{N}_s \rangle^2}{\left|\frac{d}{dg}\langle \hat{N}_s \rangle|_{g=0}\right|^2}, \qquad (9)$$

which states that the variation of the signal must be comparable to the noise of the output intensity.

For the ideal seeded configuration, where no losses are present, the sensitivity is (see Methods for derivation):

$$g_{min}^2 = \frac{1}{4\cosh^2(2G)(1+|a_i|^2)}, \qquad (10)$$

which is similar to the minimum detectable phase of the standard SU(1,1) interferometer,[34,35] and indicates the detection of a single output photon during the finite measurement time.

Both internal loss (between the two OPAs) and external/detection loss (after the interferometer) affect the sensitivity of measurement, although in a different manner. External loss takes place after the nonlinear interference and does not affect the squeezing, thereby reducing the measured signal by a loss factor $|r_{ext}|^2$, identical to the effect of losses on classical light.[36,37] Internal loss, on the other hand, hinders the quantum correlations between the signal and idler, effectively diminishing the squeezing,[38,39] which in-turn degrades the contrast of the nonlinear interference and elevates the dark fringe level (along with its associated noise), resulting in a lower detection sensitivity.

Let us calculate the average photon number and the noise associated with the dark fringe (background) due to internal losses for exactly crossed OPAs. We use the standard modeling of loss as a beam-splitter (BS) placed inside the interferometer,[40] where vacuum may enter through the unused port of the BS. We apply the BS loss to the ideal squeezing-enhanced Raman scheme, seeded with a strong (classical) coherent state $|a_i\rangle$ for the idler, and obtain for the resonant signal:

$$\langle \hat{N}_s^{(4)} \rangle = |t|^2(1+|a_i|^2)\sinh^2(g_r)\cosh^2(2G) + |r|^2\sinh^2(G), \qquad (11)$$

where $r$ represents the loss and $t$ the transmission inside the interferometer ($|r|^2 + |t|^2 = 1$). Equation (11) shows two contributions to the measured signal at the output: the resonant Raman signal (left), which is similar to the lossless case, is reduced by the transmission coefficient $|t|^2$, and the loss term (right, proportional to $|r|^2$) which corresponds to amplification of the vacuum by OPA2, and causes direct elevation of the dark fringe level. Since this term does not depend on the Raman sample, it limits the sensitivity of the measurement. The noise associated with the background due to loss is:

$$\sigma_{loss} = \sqrt{\langle \hat{N}_s^2 \rangle_{loss} - \langle \hat{N}_s \rangle_{loss}^2} \qquad (12)$$
$$= \tfrac{1}{2}|r|^2\sinh(2G).$$

Equation (12) represents the background noise of the dark fringe, which limits the ability to detect the small Raman signal. Note that plugging Eq. (12) and the derivative of Eq. (11) into Eq. (9) results in a diverging expression for the sensitivity $g_{min}^2$ for exactly crossed amplifiers ($\phi_r = \pi$). Thus, the optimal working point of the interferometer (defined as the relative phase between the OPAs where the sensitivity is optimal) may vary with the internal loss. This behavior is shown in Fig. 3a, which displays $g_{min}^2$ as a function of the phase of OPA2 ($\phi$, where $\phi_r = \pi + \phi$) for various loss values. For no loss, the optimal working point (defined as the value of $\phi$ which bestows a minimum on $g_{min}^2$), is $\phi_{opt} = 0$, as expected. Once loss is introduced, $\phi_{opt}$ increases due to the dark fringe noise. However, the squeezing enhancement still improves the minimum detectable gain below the shot-noise limit even in the presence of considerable losses, as long as the phase of OPA2 is tuned appropriately. Figure 3b introduces coherent seeding of the idler port for various intensities (number of photons) with constant 20% loss. The stimulation partially compensates for the effect of loss, improving the sensitivity plot in two ways: First, It pushes the optimal phase $\phi_{opt}$ towards the ideal case $\phi_{opt} = 0$, and also reduces the minimum detectable gain $g_{min}^2$. This improvement occurs because the seeding also stimulates the external OPAs to generate more signal-idler photon pairs, thereby increasing the total number of correlated photons that probe the sample.

Note that in our analysis, we mostly focused on the interferometer itself and on the way it interacts with the Raman sample, using the most straightforward detection—direct intensity measurements at the signal port. It is also important to consider other detection methods, such as: parity detection[41,42] which is ideal for loss-free schemes, homodyne measurement[43] which achieves higher sensitivity for the seeded interferometer, parametric homodyne[23] which is optically broadband and robust to detection losses, or a truncated interferometer[39,44] which may be preferred if the parametric amplifiers are lossy. In the Methods, we also discuss measurement at the idler port using lock-in detection (squeezing-enhanced version of SRS). The calculations can also be carried out using Fisher information analysis,[45,46] which extracts the phase from positive-operator valued measure, rather than through the error-propagation method we discussed here.

## DISCUSSION
We presented a new method for Raman spectroscopy, which utilizes the squeezed light inside a nonlinear interferometer to enhance the resonant Raman signal and to suppress the non-resonant background in the most general Raman sample. This configuration can be considered as a "Raman black box" with a squeezing-enhanced signal that is compatible with any standard Raman technique—spontaneous or stimulated, CARS or SRS. Being a "black box", it is applicable in combination with any other classical interference scheme that may be used to enhance the detection or to filter out noise as is the case in many variations of Raman spectroscopy.[47] The generality of this squeezing-enhancement along with its resilience to loss and to experimental errors, deem it highly applicable for any field where spectroscopic detection of trace-chemicals is needed.[48]

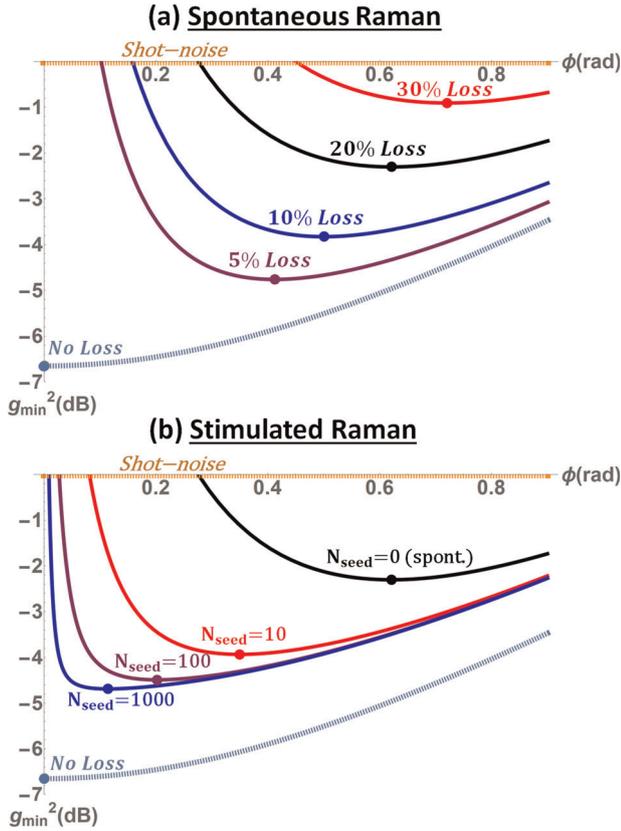

**Fig. 3** Sensitivity of the squeezing-enhanced Raman scheme compared to the shot-noise limit, for two different signal-idler inputs: **a** vacuum, and **b** seeded by a coherent idler, where the gain of the OPAs is fixed at a moderate level of $G = 0.7$ (6.65 dB of squeezing). The minimum detectable Raman gain in the sample (relative to the shot-noise limit) is shown for various configurations of seeding and internal loss as a function of the phase of OPA2 (at $\phi = 0$ the OPAs are exactly crossed). For each curve, the points highlighted with a circle represent the ideal working point of the interferometer. **a** Shows the sensitivity of the unseeded scheme relative to spontaneous Raman spectroscopy (0 dB) for various values of internal loss, indicating that sub-shot-noise sensitivity can be obtained with practical levels of internal loss (up to 30%). **b** shows the sensitivity of a Raman interferometer with fixed internal loss (20%) when seeded with various intensities of a coherent idler ($|a_i|^2 = 0, 10, 100, 1000$). Note that all sensitivities are shown relative to the shot-noise limit at their specific seeding level ($1/|a_i|^2$). Clearly the seeded configuration maintains substantial reduction below the shot-noise limit with practical levels of internal loss (and showing an improvement over the unseeded case)

## METHODS

In the following section, we first provide a short overview of the two-mode quadratures discussed in the text. We then review the SU(1,1) interferometer, and derive its squeezing-enhancement of the measurement of a linear optical phase. We then derive in more detail the major quantum expressions presented in the article: the average photon number at the signal output and its fluctuations (noise), the minimum detectable gain with coherent seeding of the interferometer, and with internal loss. To analyze the robustness of the proposed Raman interferometer to experimental imperfections, we provide additional results for non-ideal configurations, such as unbalanced interferometric detection or inaccurate orientation (phase) of the Raman sample.

### Two-mode quadratures

The basic description of the signal-idler oscillation consists of two fields with frequencies $\omega_{s,i}$ where $\Omega = \frac{\omega_s + \omega_i}{2}$ is the center frequency (and, in FWM, the frequency of the pump) and $\omega = \frac{\omega_s - \omega_i}{2}$ is the frequency separation. We start from the electric field operator of a two-mode (signal and idler) oscillation:

$$E(t) = a_s e^{-i\omega_s t} + a_i e^{-i\omega_i t} + c.c, \quad (13)$$

where $a_{s,i}$ are the annihilation operators of the signal/idler modes at frequencies $\omega_{s,i}$. The standard procedure is to rewrite the electric field in single-mode quadrature components using the standard definition of single-mode quadrature operators for the signal/idler modes $X_{s,i} = a_{s,i} + a_{s,i}^\dagger$ and $Y_{s,i} = i(a_{s,i}^\dagger - a_{s,i})$, which defines the quadratures of each mode, yielding

$$E(t) = X_s \cos(\omega_s t) + Y_s \sin(\omega_s t) \\ + X_i \cos(\omega_i t) + Y_i \sin(\omega_i t), \quad (14)$$

Since the signal and idler are generated symmetrically around the degenerate carrier frequency $\Omega$ (the pump frequency in FWM), their frequencies can be written as $\omega_s = \Omega + \omega$ and $\omega_i = \Omega - \omega$, which allows to rearrange the electric field of Eq. (14), using simple trigonometric manipulations:

$$E(t) = \tfrac{1}{2}[(X_s + X_i)\cos(\omega t) + (Y_s - Y_i)\sin(\omega t)]\cos(\Omega t) \\ + \tfrac{1}{2}[(Y_s + Y_i)\cos(\omega t) + (X_i - X_s)\sin(\omega t)]\sin(\Omega t) \quad (15)$$

We can now identify:

$$X_{2M}(t) \equiv (X_s + X_i)\cos(\omega t) + (Y_s - Y_i)\sin(\omega t), \quad (16)$$

$$Y_{2M}(t) \equiv (Y_s + Y_i)\cos(\omega t) + (X_i - X_s)\sin(\omega t), \quad (17)$$

as the cosine and sine quadrature components of the combined two-mode electric field, but now with respect to the common local oscillator at the carrier frequency $\Omega$, just like the standard case of a single-mode field. Specifically, when the intensity difference $(X_i - X_s)$ and phase sum $(Y_s + Y_i)$ are squeezed, this is equivalent squeezing of one two-mode quadrature, just like single-mode squeezing with the same squeezing ellipse. The only difference compared to the standard degenerate squeezing is that the two-mode quadratures in this definition are explicitly time-dependent.

We can now re-insert the single-mode quadratures $X_{s,i} = a_{s,i} + a_{s,i}^\dagger$ and $Y_{s,i} = i(a_{s,i}^\dagger - a_{s,i})$ into Eqs (16) and (17) to obtain:

$$X_{2M}(t) = (a_s + a_i^\dagger)e^{-i\omega t} + c.c, \quad (18)$$

$$Y_{2M}(t) = i(a_s^\dagger - a_i)e^{-i\omega t} + c.c, \quad (19)$$

which directly leads us to the definition of $X = a_s + a_i^\dagger$ and $Y = i(a_s^\dagger - a_i)$. This definition is elegant because it describes within one unified framework both two-mode squeezing and single-mode squeezing and transfers intact the intuitive picture of squeezing of the quadrature map from the single mode onto the two-mode case. Mathematically, the two-mode quadratures obey the standard commutation relation $[\hat{X}, \hat{Y}] = 2i$ just like single-mode quadratures, and evolve in a parametric amplifier just like the standard squeezing operation, with $\hat{X}_{out} \rightarrow \hat{X}_{in} e^{G}$, $\hat{Y}_{out} \rightarrow \hat{Y}_{in} e^{-G}$ (where G is the squeezing ratio). Although the two-mode quadratures are non-Hermitian $X \neq X^\dagger$, the two-mode quadrature is a measurable quantity, since it commutes with its Hermitian conjugate $[X, X^\dagger] = 0$ indicating that both $\text{Re}\{X\} = X + X^\dagger = (X_s + X_i)$ and $\text{Im}\{X\} = i(X^\dagger - X) = (Y_s - Y_i)$ are Hermitian and can be measured simultaneously to provide the complete quadrature information. A more expanded study about this two-mode terminology along with its detection is presented in.[23]

### Review of the SU(1,1) interferometer

Interferometric measurements allow for highly sensitive detection of any physical phenomenon that induces an optical phase shift. The phase sensitivity of an interferometric scheme depends on both the illumination source and the configuration of the interferometer: Standard SU(2) interferometers, such as the Michelson or Mach-Zehnder, achieve a phase sensitivity $(\theta_{min}^2)$ of $1/N$—the shot-noise limit, when fed with coherent light ($N$—the average number of photons that traversed the interferometer during the detection time. SU(2) interferometers can surpass this limit when the unused port of the interferometer is fed with squeezed light. The SU(1,1) nonlinear interferometer is a fundamentally different type of interferometric detector, where nonlinear gain media (OPAs) replace the beam splitters, and squeezed light is generated within the interferometer itself without the need to feed externally. Additionally, the SU(1,1) interferometer can be robust to detection losses.[49]

In a parametric process, the direction of energy transfer - from the pump to the signal and idler or vice-versa—depends on the relative phase between the pump and a signal-idler pair, such that either amplification ($\phi_r = 0$) or attenuation ($\phi_r = \pi$) of the signal-idler pair occurs. The two OPAs of an SU(1,1) interferometer are arranged in series with opposite phase, where the attenuation axis of OPA2 matches the amplification axis OPA1 and vice-versa (setting $\phi_r = 0$ in OPA1 and $\phi_r = \pi$ in OPA2). Thus, if the gain of both OPAs is equal, the output quantum state of the light remains unchanged from the input (unless the phase of the light is altered between the amplifiers). Indeed, when the standard SU(1,1) interferometer measures a linear phase shift $\theta$ between the two OPAs (assuming vacuum input for now), the number of signal (or idler) photons at the output is given by:

$$\langle \hat{N}_s \rangle_{SU(1,1)} = \sinh^2(2G)\sin^2\left(\frac{\theta}{2}\right), \quad (20)$$

where $G$ is the gain of the OPAs. The second moment is:

$$\langle \hat{N}_s^2 \rangle_{SU(1,1)} = \sinh^4(2G)\sin^4\left(\frac{\theta}{2}\right) + \sinh^2(2G)\sin^2\left(\frac{\theta}{2}\right). \quad (21)$$

If no phase shift is present, the output is vacuum (identical to the input), with $(\langle \hat{N}_s \rangle, \langle \hat{N}_s^2 \rangle = 0)$, allowing signal detection which is background-free. The phase sensitivity, obtained by error-propagation analysis (see Eq. 28 for additional details), is given by:

$$\theta^2_{min} = \frac{\langle \hat{N}_s^2 \rangle - \langle \hat{N}_s \rangle^2}{\left|\frac{d}{d\theta}\langle \hat{N}_s \rangle|_{\theta=0}\right|^2}, \quad (22)$$
$$= \frac{1}{\sinh^2(2G)} \approx \frac{1}{N_{sq}^2},$$

where $N_{sq}$ is the number of squeezed signal and idler photons generated inside the interferometer. The result of Eq. (22) shows sub-shot-noise scaling, allowing for super sensitive phase detection (ideally Heisenberg sensitivity of $1/N_{sq}^2$).

### Ideal squeezing-enhanced configuration

We start by expressing the output field operators (creation and annihilation) of the signal and idler as a function of the input field operators in a three-stage propagation: through OPA1 (optical parametric amplifier), through the Raman sample, and last through OPA2. Assuming an undepleted (classical) pump, the signal-idler field operators evolve in an OPA as[50]

$$\hat{a}^{(1)}_{s,i} = \left(\cosh(G) + i\frac{\Delta q z}{2G}\sinh(G)\right)\hat{a}^{(0)}_{s,i} + \frac{\gamma l_p z}{G}\sinh(G)\hat{a}^{\dagger(0)}_{i,s}, \quad (23)$$

where $G$ is the gain of the OPA, $\Delta q = \Delta k + 2\gamma l_p$ is the total phase mismatch, comprised of the bare phase mismatch $\Delta k = 2k_p - k_i - k_s$ and $2\gamma l_p$ the nonlinear phase induced by the pump ($l_p$ is the pump intensity and $\gamma$ is the Kerr nonlinear coefficient), $z$ is the length of the medium, and $\hat{a}^{(0)}_{s,i}, \hat{a}^{\dagger(0)}_{s,i}$ are the field operators before the OPA. The gain is given by $G = z\sqrt{\gamma^2 l_p^2 - \frac{\Delta q^2}{4}}$. For simplicity, we assume perfect phase matching, $\Delta q = 0$. This assumption is not critical for the calculation, which can be carried out just as well with any phase mismatch, and served only for simplification of the final expression. From a practical point of view, OPA1 and OPA2 can be phase matched across a broad spectrum with careful designing of the nonlinear medium, while phase matching in the Raman medium depends on the nonlinear properties of the sample itself. We may now define the parameteres $A \equiv \cosh(G)$, $B \equiv \sinh(G)$. Using Eq. (23), we can express the field operators at the output of OPA1, the Raman sample and OPA2 with the field operators at the input. OPA1 performs amplification at $\phi_r = 0$, such that:

$$\hat{a}^{(1)}_{s,i} = A\hat{a}^{(0)}_{s,i} + B\hat{a}^{\dagger(0)}_{i,s}. \quad (24)$$

The Raman sample then follows up on the amplification of OPA1, but at a relative phase of $\varphi_r = \frac{\pi}{2}$ and different gain parameters ($g_r$, $C$, $D$). Thus, we apply a phase of $\pi/4$ to both the signal and the idler as we propagate the operators through the sample:

$$\hat{a}^{(2)}_{s,i} = C\hat{a}^{(1)}_{s,i}e^{i\pi/4} + D\hat{a}^{\dagger(1)}_{i,s}e^{-i\pi/4}. \quad (25)$$

Finally, OPA2 is set orthogonal to OPA1 (a total relative phase of $\phi_r = \pi$), and accordingly we once more apply a $\pi/4$ to the signal and idler:

$$\hat{a}^{(3)}_{s,i} = A\hat{a}^{(2)}_{s,i}e^{i\pi/4} + B\hat{a}^{\dagger(2)}_{i,s}e^{-i\pi/4}. \quad (26)$$

The photon number at the final output can be calculated using the relation between the operators before and after interferometer, while applying $\left\langle 0_s,0_i\left|\hat{a}^{\dagger(0)}_{s,i}\hat{a}^{(0)}_{s,i}\right|0_s,0_i\right\rangle = 0$ and $[\hat{a}^{(0)}_{s,i},\hat{a}^{\dagger(0)}_{s,i}] = 1$, resulting in:

$$\begin{aligned}\langle \hat{N}_s^{(3)}\rangle &= \langle 0_s,0_i|\hat{a}^{\dagger(3)}_s\hat{a}^{(3)}_s|0_s,0_i\rangle \\ &= |D|^2(|A|^2 + |B|^2 + 2|A||B|) \\ &= \sinh^2(g_r)\cosh^2(2G),\end{aligned} \quad (27)$$

The minimum detectable Raman gain is calculated using an error-propagation analysis, demanding that

$$g^2_{min} = \frac{\langle \hat{N}_s^2 \rangle - \langle \hat{N}_s \rangle^2}{\left|\frac{d}{dg_r}\langle \hat{N}_s \rangle|_{g_r=0}\right|^2} \quad (28)$$

where the numerator represents the background noise and the denominator represents the relative change of the signal intensity as the Raman gain is varied. For the ideal, lossless case, we obtain:

$$\sqrt{\langle \hat{N}_s^2 \rangle - \langle \hat{N}_s \rangle^2} = |C||D|(|A|^2 + |B|^2), \quad (29)$$

$$\frac{d}{dg_r}\langle \hat{N}_s \rangle = 2|C||D|(|A|^2 + |B|^2)^2, \quad (30)$$

both terms approach 0 at $g_r = 0$ ($D = 0$, $C = 1$) but their ratio is finite. taking the limit of $g_r \to 0$, we obtain:

$$g^2_{min} = \frac{1}{4\cosh^2(2G)}. \quad (31)$$

### Seeded configuration

The calculation shown in Eq. (31) can be generalized for the seeded configuration. In the main text, we present the result for a coherent idler seed $|\alpha_i\rangle$ used as the input to OPA1. The calculation for this case is conceptually the same as the unseeded case, but applied to a different input. The average of the photon number requires no additional assumptions, and is given by:

$$\begin{aligned}\langle \hat{N}_s^{(3)}\rangle &= \left\langle 0_s,\alpha_i|\hat{a}^{\dagger(3)}_s\hat{a}^{(3)}_s|0_s,\alpha_i\right\rangle \\ &= (1 + |\alpha_i|^2)\sinh^2(g_r)\cosh^2(2G),\end{aligned} \quad (32)$$

which forms a squeezing-enhanced signal for CARS. Alternatively, the average photon number at the seeded (idler) port is given by:

$$\begin{aligned}\langle \hat{N}_i^{(3)}\rangle &= \tfrac{1}{4}\big[3|\alpha_i|^2 - 1 + (|\alpha_i|^2 + 1)\\ &\cdot(2\cosh^2(2G)\cosh^2(2g_r) - \cosh(4G))\big],\end{aligned} \quad (33)$$

where Eq. (33) shows the amplification of the idler. In SRS, the net amplification of the idler is measured via lock-in detection. The expression for SRS is then:

$$\begin{aligned}\langle \hat{N}_i^{(3)}\rangle - |\alpha_i|^2 &= \tfrac{1}{2}(1 + |\alpha_i|^2)\sinh^2(g_r)(1 + \cosh(4G)) \\ &= (1 + |\alpha_i|^2)\cosh^2(2G)\sinh^2(g_r).\end{aligned} \quad (34)$$

Following the same steps of Eq. (31), the minimum detectable gain for the seeded case is:

$$g^2_{min} = \frac{1}{4\cosh^2(2G)(1 + |\alpha_i|^2)}. \quad (35)$$

### Robustness to loss, unbalanced detection and phase inaccuracy

Loss between the two OPAs can be treated as a beam-splitter, which mixes the field operators $\hat{a}^{(2)}_{s,i}$ with an additional vacuum mode $\hat{b}^{(0)}_{s,i}$, indicating that the operators at the entrance of OPA2 are $\hat{a}^{(3)}_{s,i} = t\hat{a}^{(2)}_{s,i} \pm r\hat{b}^{(0)}_{s,i}$, where $t$, $r$ are the amplitude transmission and reflection coefficients of the assumed beam-splitter. The number of photons is given by:

$$\langle \hat{N}_s^{(4)} \rangle = |t|^2\sinh^2(g_r)\cosh^2(2G) + |r|^2\sinh^2(G), \quad (36)$$

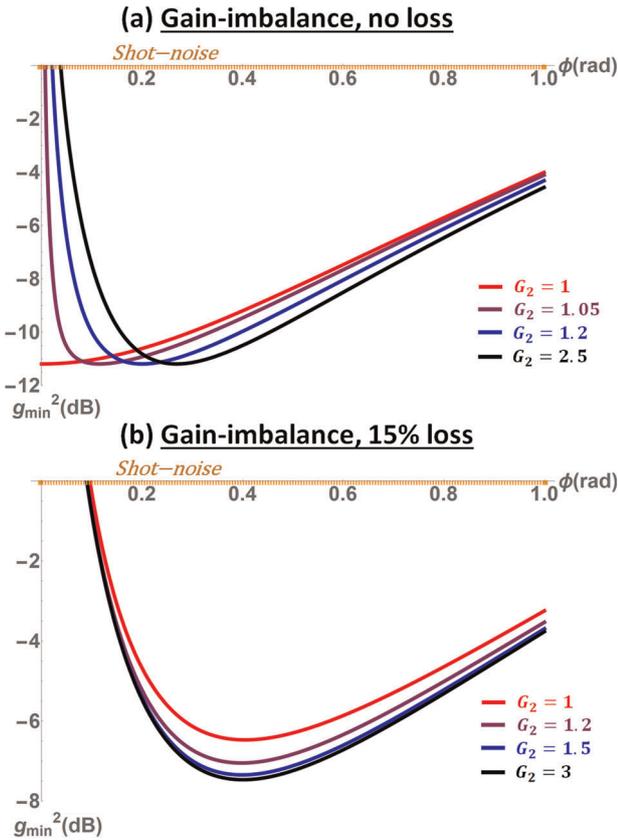

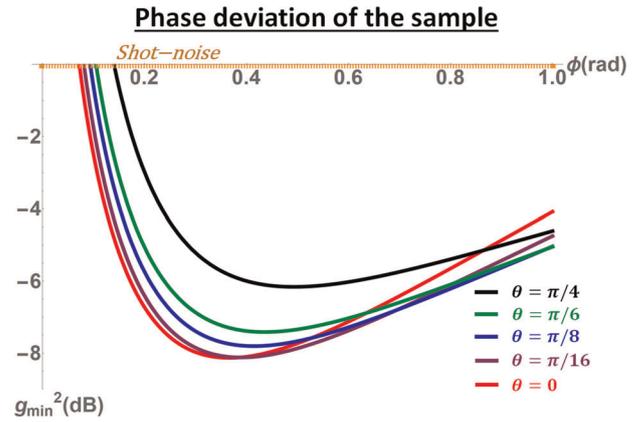

**Fig. 4** **a** Sensitivity of the squeezing-enhanced Raman scheme compared to the shot-noise limit for the unbalanced configuration, calculated with $G_1 = 1$ and no loss ($|r|^2 = 0$). **b** Sensitivity of the unbalanced configuration with internal loss of 15% ($|r|^2 = 0.15$)

**Fig. 5** Robustness to imprecision of the optical phase of the Raman sample: The sensitivity of the squeezing-enhanced Raman scheme compared to the shot-noise limit for the balanced configuration, when the phase of the Raman sample deviates from $\phi_r = \pi/2$ for $\theta > 0$. The interferometer is set to be balanced with $G_1 = G_2 = 1$ and internal loss is set to $|r|^2 = 0.15$. The sensitivity remains close to that of the ideal case for small deviations (up to $\pi/8$)

and the background noise is:

$$\sqrt{\langle \hat{N}_s^2 \rangle_{loss} - \langle \hat{N}_s \rangle_{loss}^2} = \frac{1}{2}|r|^2 \sinh(2G). \quad (37)$$

Plugging Eq. (37) into Eq. (28) results in a diverging limit at $g_r \to 0$. Thus, the relative phase before OPA2 must also be taken into consideration when calculating the sensitivity for the squeezing-enhanced Raman scheme with internal loss. Analytical expressions are no longer simple and intuitive enough to be provided in this paper, so it is best to refer to the numerical plots of Fig. 3 provided in the main text.

The scheme is also effective in non-symmetric conditions, such as unbalanced interferometric detection ($G_1 \neq G_2$):[51,52] This can occur because of different pump intensities before the different OPAs, or due to unsuppressed non-resonant background. This behavior is shown in Fig. 4a for the lossless case, where the sensitivity remains identical to the balanced case, but the optimal working point of the interferometer shifts in phase to compensate for the added noise at $\phi = 0$. If internal loss is introduced, major unbalancing of the OPAs can somewhat compensate for the loss and improve the sensitivity, as shown in Fig. 4b, and also discussed in.[37,53] The unbalanced interferometer boasts other potential advantages, such as high-bandwidth two-mode homodyne detection. This is achieved by setting $G_2 > G_1$, where the main idea is that for a sufficiently strong difference in gain, the light output of OPA2 is proportional to the quadrature it amplifies, using the pump as the local oscillator.[23]

It is also important to consider the amplification angle of the Raman sample, which ideally would be at the mid-point between the amplifiers, as previously discussed in Fig. 2. We now discuss the non-ideal case, where the relative phase of the light at the Raman sample deviates from $\phi_r = \pi/2$ by some adverse phase $\theta$. This phase arises, for example, from inaccurate compensation of linear dispersion inside the interferometer. In Fig. 5, we show the sensitivity for different values of the phase deviation, where for relatively small phase deviations of the Raman sample, the sensitivity is nearly unaffected by this error. Note that we have assumed here that the sample has a pure resonant response, and in general, this behavior depends on the exact ratio between the resonant and non-resonant gains.

### DATA AVAILABILITY
All the data and calculations that support the findings of this study are available from the corresponding author upon reasonable request.

### CODE AVAILABILITY
The code used to calculate the results given in this paper will be made available from the corresponding author upon reasonable request.

Received: 14 April 2019; Accepted: 9 September 2019;
Published online: 01 October 2019

## ACKNOWLEDGEMENTS
This research was supported by grant No.44/14 from the Israel Science Foundation (FIRST program for high-risk high-gain research). We thank Dr. Emanuele Dalla Torre and Prof. Richard Berkovits for their thoughtful comments. We are grateful to Jose Luis Gomez-Munoz and Francisco Delgado for their "Quantum notation" add-on to the Mathematica software, which we have used for the quantum calculations in this work.

## AUTHOR CONTRIBUTIONS
Y.M. proposed the main concept and carried out the calculations, L.B. and M.R. participated in the analysis of results and continuous discussions, Y.M. and A.P. developed the theoretical framework. All authors contributed to writing of the paper.

## COMPETING INTERESTS
The authors declare no competing interests.

## ADDITIONAL INFORMATION
**Correspondence** and requests for materials should be addressed to A.P.

**Reprints and permission information** is available at http://www.nature.com/reprints.

**Publisher's note** Springer Nature remains neutral with regard to jurisdictional claims in published maps and institutional affiliations.